\documentclass[preprint,12pt]{elsarticle}

\usepackage{amssymb}

\usepackage{float}
\usepackage{amstext}
\usepackage{graphicx}
\usepackage{esint}

\newcounter{bla}

\journal{Computer Physics Communications}

\begin{document}

\begin{frontmatter}



\title{ZMCintegral-v5.1: Support for Multi-function Integrations on GPUs}


\author[a]{Xiao-Yan Cao}

\author[b]{Jun-Jie Zhang\corref{author}}

\cortext[author]{Corresponding author.\\
\textit{E-mail address:} zjacob@mail.ustc.edu.cn(Jun-Jie Zhang)}

\address[a]{Xi'an Research Institute of High-Tech}

\address[b]{Northwest Institute of Nuclear Technology}


\begin{abstract}
In this new version of ZMCintegral, we have added the functionality
of multi-function integrations, i.e. the ability to integrate more
than $10^{3}$ different functions on GPUs. The Python API remains
the similar as the previous versions. For integrands less than 5 dimensions,
it usually takes less than 10 minutes to finish the evaluation of
$10^{3}$ integrations on one Tesla v100 card. The performance scales
linearly with the increasing of the GPUs.
\end{abstract}

\end{frontmatter}


\noindent
{\bf NEW VERSION PROGRAM SUMMARY}

\begin{small}
\noindent
{\em Program Title: } ZMCintegral \\                                          
{\em Licensing provisions: } Apache-2.0 \\
{\em Programming language: } Python \\
{\em Journal reference of previous version: } Hong-Zhong Wu, Jun-Jie Zhang, Long-Gang Pang, Qun Wang, Comput. Phys. Commun. 248(2020):106962 and Jun-Jie Zhang, Hong-Zhong Wu, Comput. Phys. Commun. 251(2020):107240\\
{\em Does the new version supersede the previous version?: } Yes \\
{\em Reasons for the new version: } When solving the Boltzmann
equation with radiations\cite{Oxenius1986}, one encounters different
collision integrals for different energy beams. In the relativistic
QED plasma, the collision terms involve various Feynman graphs\cite{Morozov2009}
and usually the contribution from each graph is of great interest.
In these circumstances one needs to integrate many functions of different
forms simultaneously. In our previous versions\cite{Wu2019,Zhang2020},
we focused on single integration of high dimensions and functions with
parameters. Therefore, it is necessary to include the functionality for integrating many functions which have different dimensions, forms and integration domains.\\
{\em Summary of revisions:}\\
 
\begin{itemize}
\item Multi-function Integrations \\
 
\end{itemize}
\noindent Suppose we have a series of integrations defined as
\begin{eqnarray}
f_{n}(\mathbf{x}) & = & a_{n}\text{cos}(\mathbf{k}_{n}\cdot\mathbf{x})+b_{n}\text{sin}(\mathbf{k}_{n}\cdot\mathbf{x}),\label{eq:multi-integration}
\end{eqnarray}
where $n=1,2,3,...,100$. The above integration series can be treated
as a set of Harmonic bases if one wishes to evaluate the contribution
of each Harmonic mode. In our previous versions, these series cannot
be manipulated in a convenient and efficient way. It is worth noting
that the integration domains or dimensions can be different, for example

\noindent 
\begin{eqnarray}
g_{n}(x_{1},x_{2}) & = & a_{n}|x_{1}+x_{2}|\ \text{for }0<n<50\nonumber \\
g_{n}(x_{1},x_{2},x_{3}) & = & b_{n}|x_{1}+x_{2}-x_{3}|\ \text{for }50\leq n\leq100.\label{eq:diff}
\end{eqnarray}
The support of multi-functions gives the users full flexibility to
integrate as many different functions as possible.
\begin{itemize}
\item Test on GPUs
\end{itemize}
As an illustrative example, we report the solution of Eq. (\ref{eq:multi-integration})
with $\mathbf{x}=(x_{1},x_{2},x_{3},x_{4})$, $a_{n}=b_{n}=1$. The
ranges for all components are taken to be $[0,1]$ and $\mathbf{k}_{n}=(\frac{n+50}{2\pi},\frac{n+50}{2\pi},\frac{n+50}{2\pi},\frac{n+50}{2\pi})$
such that the integration is highly fluctuating around the zero line.
The hardware condition in this case is taken to be: Intel(R) Xeon(R)
Silver 4110 CPU@2.10GHz CPU with 10 processors + one Nvidia Tesla
V100 GPU. The results are shown in Fig. \ref{tab:Solution-of-Eq.}. 

\begin{figure}
\begin{centering}
\includegraphics[scale=0.6]{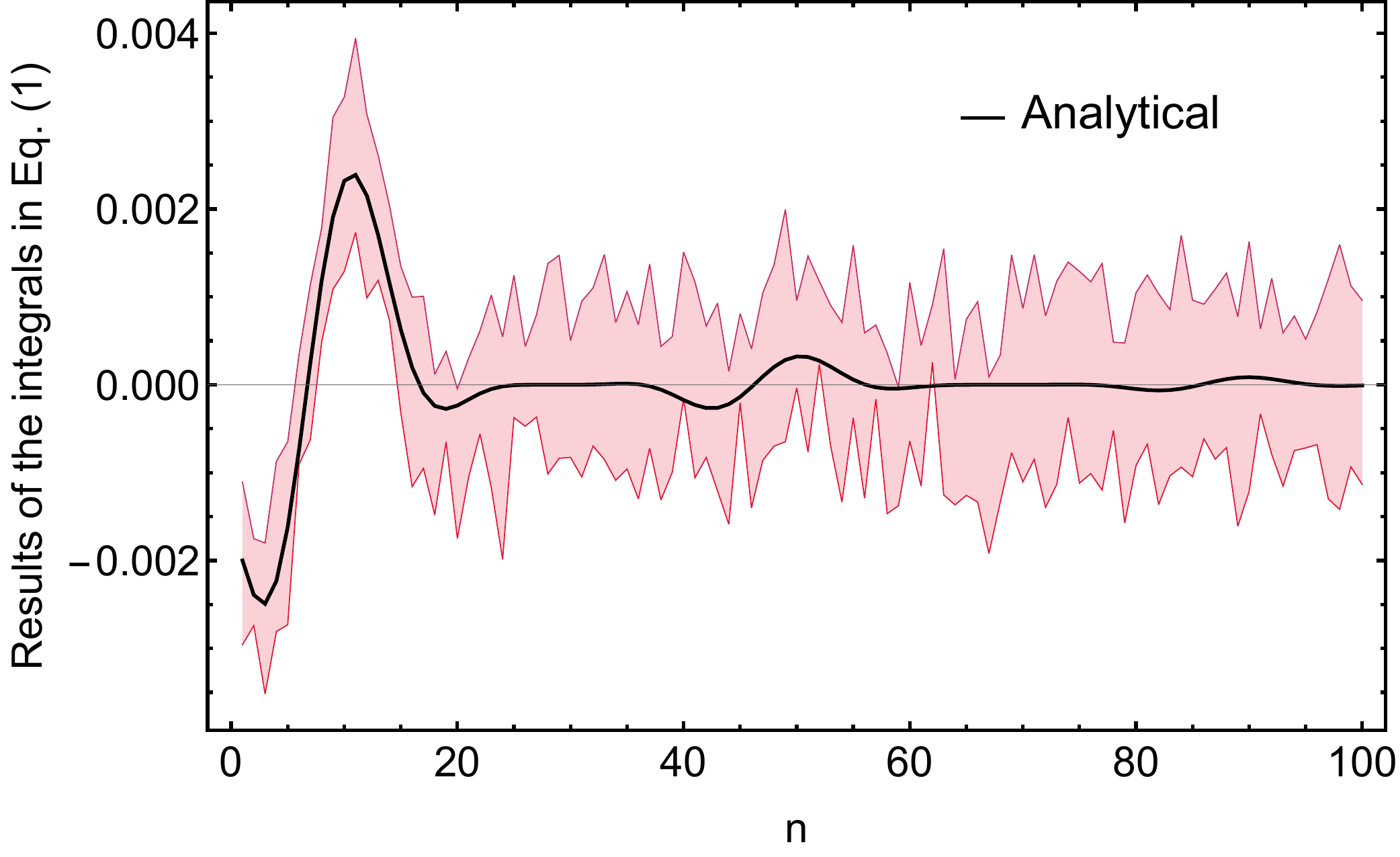}
\par\end{centering}
\caption{\label{tab:Solution-of-Eq.}Integration results of Eq. (\ref{eq:multi-integration})
at various $n$. The red shaded area, drawn from $[\overline{F_{n}}-\triangle F_{n},\overline{F_{n}}+\triangle F_{n}]$
with $F_{n}$ denoting the result of the $n\text{-th}$ integral,
is obtained via ZMCintegral (analytical results in solid black line).
Here, $\overline{F_{n}}$ and $\triangle F_{n}$ are the average value
and standard deviation of 10 independent evaluations. It takes about
one minute for each independent calculation (of all series in Eq.
(\ref{eq:multi-integration})) on one Tesla V100 card. For each $n$
we have taken $10^{6}$ samples.}

\end{figure}

{\em Nature of problem:} ZMCintegral is an easy to use Python package
for doing high dimensional integrations on distributed GPU clusters.
Using the Python libraries Numba\cite{Kwan2015} and Ray\cite{PMoritz2018},
as well as the NVIDIA CUDA\cite{cuda01} capability, ZMCintegral
offers a succinct Python interface to evaluate numerical integrations
for physical problems. In this updated version, we mainly focus on
the problems where the users have various functions to integrate. These
integrations can take different forms and domains.\\
 {\em Solution method:} This new version contains three Python
classes. ZMCintegral\_normal utilizes the stratified-sampling and
heuristic-tree-search techniques, while ZMCintegral\_functional and
ZMCintegral\_multifunctions use the direct-Monte Carlo method for
each integrand which benefits mainly from the heavily distributed
GPU clusters.\\
 {\em Additional comments:} If the integrations are high-dimensional
(e.g. dimensionality of 8-12), users are encouraged to use ZMCintegral\_normal.
If the integrations are middle-dimensional (e.g. dimensionality of
1-7) but with a large parameter space, we suggest users to try ZMCintegral\_functional.
If the integrations contains many different integrands and domains
(e.g. $10^{4}$ different integrations), then ZMCintegral\_multifunctions
is suggested. The detailed instructions can be found here: \cite{zmc_github}.
\\
 \end{small}











\section*{\textit{Acknowledgment}}

The authors are supported in part by the Major State Basic Research
Development Program (973 Program) in China under Grant No. 2015CB856902
and by the National Natural Science Foundation of China (NSFC) under
Grant No. 11535012. The Computations are performed at the GPU servers
of department of modern physics at USTC. We are thankful for the valuable
help from Prof. Qun Wang of department of modern physics at USTC.

\bibliographystyle{elsarticle-num}
\bibliography{ZMCintegral}

\end{document}